\documentclass[prd,twocolumn]{revtex4}
\usepackage{epsf}
\usepackage{epsfig,graphicx,psfrag,amsfonts,amssymb}

\usepackage[dvips]{color}

\usepackage{amsmath,amssymb}

\begin{document}

\title{Observational Constraints on Axions as  Quintessence in String Theory}
\author{Gaveshna Gupta$^1$\footnote{gaveshna.gupta@gmail.com}, Sudhakar Panda$^2$\footnote{panda@hri.res.in},  Anjan A Sen$^1$\footnote{anjan.ctp@jmi.ac.in}}
\affiliation{$^1$ Centre of Theoretical Physics, Jamia Millia Islamia, New Delhi-110025, India}
\affiliation{$^2$ Harish-Chandra Research Institute, Chhatnag Road,
Jhunsi, Allahabad-211019, India}

\date{\today}

\begin{abstract}
We study the observational constraints for the axion models in string theory which can successfully act as quintessence. The evolution of the universe in this model is sensitive to the initial value of the axion field. This initial value of the axion field controls deviation of the cosmic evolution from the $\Lambda$CDM behaviour. We use the recent Union2 Supernova Type Ia dataset as well as the data from BAO measurements, the WMAP measurement of the shift parameter and the $H(z)$ measurements. Using these data, the reconstructed equation of state of the axion field has extremely small deviation from the cosmological constant $\omega = -1$ even at $2\sigma$ confidence level. One interesting outcome of this analysis is that one can put bound on the SUSY breaking scale using cosmological measurements assuming the values of different string related parameters and vice versa. 
\end{abstract}

\maketitle

\date{\today}

\maketitle

\section{Introduction}

Cosmological observations\cite{obs1} \cite{cmb} \cite{wmap} indicate that at present our universe is in an accelerating phase of expansion. The general belief is that a mysterious form of energy, called the dark energy\cite{sami}, is responsible for driving the universe into such a late time accelerating phase. However, the nature of dark energy is not yet understood and remains a challange for us. So far, the best possible explaination for dark energy has been a constant vacuum energy dubbed as the cosmological constant. Even if models for dark energy  based on the cosmological constant have been consistent with observations, cosmological constant itself is plagued from an acute problem called the cosmological constant problem\cite{derev}. On the other hand, observational data still can accommodate a time varying vacuum energy known as quintessence. Infact, quintessence\cite{quint}  was proposed as a candidate for dark energy to provide a dynamical solution to the cosmological constant problem. The quintessence model is based on the proposal that the vacuum energy is not a constant but depends on a scalar field, called quintessence field, whose energy density slowly varies in time since the scalar field itself evolves in time. This results in making the equation of state for dark energy $\omega$ time dependent. The mechanism of a slowly varying scalar field puts the dynamics of quintessence in a similar framework for the dynamics of inflation. However, we should keep in mind that the energy scale of quintessence is of order $10^{-3}$ eV which is not only much less than the energy scale of inflation but also much less than that of supersymmetry breaking. Thus, quintessence model building faces the most important challange of constructing a potential for the scalar field which  can meet the required slow-roll criteria even if the supersymmetry breaking scale is much higher. To be more precise, we need the mass of the quintessence field to be of order of the Hubble constant at present epoch ($10^{-33}$ eV) and this should not be driven upto the higher scale of supersymmetry breaking. This is not easy to achieve given our experience of the hierarchy problem in standatrd model of particle physics. Moreover, the requirement of small mass of the quintessence field expressed in terms of the restrictions on higher dimensional opertors appearing in the effective field theory action reveals that even if we take the supersymmetry breaking scale to be of order 1 TeV, operators up to dimension 10 can contribute to the mass.  These operators are to be suppressed in the construction of a satisfactory model of quintessence. This tells us that not only the slow-roll requirement is very restrictive but also quintessence model building is sensitive to Planck scale physics. Hence it is sensible to construct a quintessence model in the framework of a UV complete theory of gravity like string theory. Also, it can happen that one needs to ensure the flatness of the potential for the entire period of evolution of the quintessence field which is a difficult task in the framework of effective field theory.These issues  were addressed recently in the context of string theory by Panda et al.\cite{axi}, hereafter called the PST model. This model is the first thoroughly constructed and controlled example of a cosmological quintessence scalar field model in string theory. The model is based on the idea of axion monodromy in Type II B string theory\cite{McAllister:2008hb,Svrcek:2006yi,Silverstein:2008sg} which was previously used for large field inflation. As opposed to earlier models of quintessence which were mostly restricted within the realm of effective scalar field theory and superfluous stringy ruminations, the present model is a significant step forward. This is mainly because of the fact that it provides a quintessence candidate, the axion field, with a symmetry that has power to suppress the otherwise deadly radiative corrections to its potential. Besides, the model takes into account the numerous possible corrections to the scalar potential and other difficulties arising from embedding the conceptual configuration into a compactified string model with moduli stabilization and supersymmetry breaking. In the next section, we briefly review the construction of PST model.

\section{The PST model of Quintessence:}

It is a well known fact that axion fields arise in models of string compactifications from both NS-NS and R-R sectors of the theory. The NS axion field  arises from the zero-mode of the NS two-form $B_2$ while the zero-modes of the RR two-form $C_2$ and four-form $C_4$ gives rise to the RR axion field. These fields have a symmetry under a constant shift of the fields. This symmetry, called the shift symmetry, is an exact symmetry to all orders in quantum loop expansion. Thus this symmetry can be broken only by non-perturbative effects and the resulting potential can be ensured to vary slowly. Moreover, such an aspect of the theory can be isolated from the effect of supersymmetry breaking. PST model uses the axion arising from the RR sector. The model is constructed from the flux compactification of Type IIB string theory\cite {Douglas:2006es,Giddings:2001yu}. The compactified manifold is a orientifolded Calabi-Yau three fold which supports the presence of three-form and five-form fluxes. Turning on the fluxes and adding some background branes produces a warped internal space. It is well known that the fluxes can stabilize the complex structure moduli fields.  The volume modulus is stabilized a la KKLT \cite{kklt}. However, for the axionic quintessence model of PST, it is desirable to generate the non-perturbative potential via gaugino condensation, required for stabilizing the volume modulus, by embedding a stack of D7-branes at the base of throat, instead of an Euclidean D3-brane generating a potential through instantons.These D7-branes wrap a four-cycle in the Calabi-Yau.  One needs to break the shift symmetry of the axion field to generate a potential for the axion. The usual method, exploiting the instanton effects, generates a periodic potential which does not lead to an acceptable quintessence model unless we allow a large number of axions to evolve coherently. PST model exploited the fact that the shift symmetry can be broken in presence of branes which are placed in highly warped regions or throats, of the compact space. The resulting axion potential is no more periodic but turns out to be approximately linear in the field i.e. linear for large values of the field. The symmetry breaking is achieved by adding a pair  of throats to the parent Calabi Yau manifold with two nontrivial two-cycles descending into each of the throats. A NS5-brane is then placed at the bottom of the first throat, where the warp factor takes the minimum value. The brane wraps a combination of the two cycles which is invariant under the orientifold symmetry. In  presence of the axion field the NS5-brane acquires  charge and tension corresponding to a D3-branes which fill up the non-compact space-time but are point-like in the compact space\cite{McAllister:2008hb}. The emergence of D3 charge for a NS5-brane, in presence of $C_2$ field, is best understood from WZ coupling term $\int C_2 \wedge C_4$.  For the purpose of  charge cancellation, an anti-NS5-brane, (i.e. NS5-brane wrapping the two-cycles with opposite orientation) is placed at the bottom of the second throat which induces charge and tension associated with anti-D3-brane.   Thus, the resulting configuration space is non-supersymmetric.  Hence, one needs to include the effect on the axion potential arising due to supersymmetry breaking besides other effects like contributions from moduli stabilization and additional warping due to extra 3-brane charges arising due to presence of axions. All these effects have been carefully analysed and resulting axion potential turns out to be approximately linear provided there exists a $Z_2$ symmetry in the Calabi-Yau manifold which interchanges the two throats containing the 5-brane and the anti-5-brane. The reader is reffered to the original paper \cite{axi} for the detailed analysis. We present below  some of the results obtained from this analysis. Taking the axion field coming from the zero mode of $C_2$ on $T^2$ spanned by the first two internal directions, $a= C_{12}$, the four dimensional axion action, neglecting the dependence of other moduli, obtained from the dimenensional reduction of the ten dimensional effective action takes the form:
\begin{equation}
S~=~\int d^4x \sqrt{-g_4} \left[ \frac{M_{pl}^2}{2} R ~ - \frac{f_a^2}{2} (\partial a)^2 \right]
\end{equation}
where the four-dimensional Planck scale, the axion decay constant and the volume of the internal space are expressed in terms of the dimensionless modulus $L$ and string coupling constant $g_s$ respectively as 
\begin{eqnarray}
M_{pl}^2 &=&  \frac{2 L^6}{(2\pi)^7 g_s^2 \alpha{'}} \nonumber\\
f_{a}^2 &=& \frac{g_s^2M_{pl}^2}{6 L^4}\nonumber\\
V &=& L^6 \alpha^{'3},
\end{eqnarray}

\noindent
where $V$ is the volume of the internal space. The ten-dimensional metric for the warped compactification is taken to be of the form:
\begin{equation}
ds^2~=~e^{2A(y)} dx_\mu dx^\mu + e^{-2A(y)} g_{ab} dy^a dy^b
\end{equation}
where $\mu = 0, 1, 2, 3$ corresponds to the non-compact directions; $a, b$ label  the six compact Calabi-Yau directions. The warp factor at the location of the NS5-brane is taken to be $e^{A_0}$ where $A_0$ is the minimum value of $A(y)$. In this case the DBI action for the NS5-brane, in presence of $C_2$ gives rise to a potential for the axion given by
\begin{equation}
V_0 = \frac{2 e^{4 A_0}}{(2\pi)^5 g_s^2 \alpha^{'2}} \sqrt{ L^4 + g_s^2 a^2}.
\end{equation}
The factor of 2, in the above, takes into account of the contributions from the 5-brane and anti-brane. In case when the axion field takes a large value i.e. $a >> L^2/g_s$, the potential is approximately linear:
\begin{equation}
V_0 = \frac{2 e^{4 A_0}}{(2\pi)^5 g_s \alpha^{'2}}  a.
\end{equation}
However, as mentioned above, the extra 3-brane charges arising due to presence of axions produces an additional warping which changes the overall volume of the compactification. Since, the volume has been already stabilized this change brings in an axion dependent potential energy. The leading contribution to the potential could be made to vanish by imposing the discrete symmetry in the internal space under which the two throats carrying the 5-brane and the anti-brane are interchanged. Nonetheless, a subleading contribution to the potential is non-zero and is estimated to be
\begin{equation}
V_1 = c M_{SB}^4 e^{2 A_0} \left( \frac{R^2}{{\alpha'} L^4} \right) a.
\end{equation}   
Here, $M_{SB}$ is the scale corresponding to the supersymmetry breaking, $c$ is a positive numerical constant of $O (1)$ and $R$ is the radius of the AdS-like throats in which the 5-brane and anti-branes are placed.
We should keep in mind that in the supergravity approximation, $R^2/{\alpha'} > 1$. Also $e^{A_0}$ in both $V_0$ and $V_1$ is the warp factor at the bottom of the throats. Always it has been considered that $a > 0$. This means that $V_1$ is dominant compared to $V_0$. However, we will consider the total axion potential to be of the form
\begin{equation}
V(a) = V_0 + V_1\equiv {\mu ^4} a
 \end{equation}
 where $\mu$ is the mass scale. Defining the canonically normalized axion field to be $\phi = f_a a$, the four dimensional action for axion-gravity system is given as
\begin{equation}
{\cal S} = \int d^{4}x \sqrt{-g_4} \left[ {M_{pl}^2\over 2} R - {1\over{2}}\partial_{\mu}\phi\partial^{\mu}\phi -V(\phi) \right],
\end{equation} 
where   
\begin{equation}
V(\phi) = \frac{\mu^4}{f_{a}} \phi
\end{equation}
with the mass scale $\mu$ being related to the other parameters as
\begin{equation}
\mu^{4} = \mu_{1} + \mu_{2},
\end{equation}

\noindent
where 
$\mu_{1} = \frac{2 e^{4 A_0}}{(2\pi)^5 g_s \alpha'^{2}}$ and  $\mu_{2} = c M_{SB}^4 e^{2 A_0} \left( \frac{R^2}{{\alpha' } L^4}\right)$.
 
 \vspace{2mm}
In the next section we  discuss the dynamics of the axion field using the above action.

\section{QUINTESSENCE FIELD DYNAMICS}

\begin{figure}[t]
\centerline{\epsfxsize=3.5truein\epsfbox{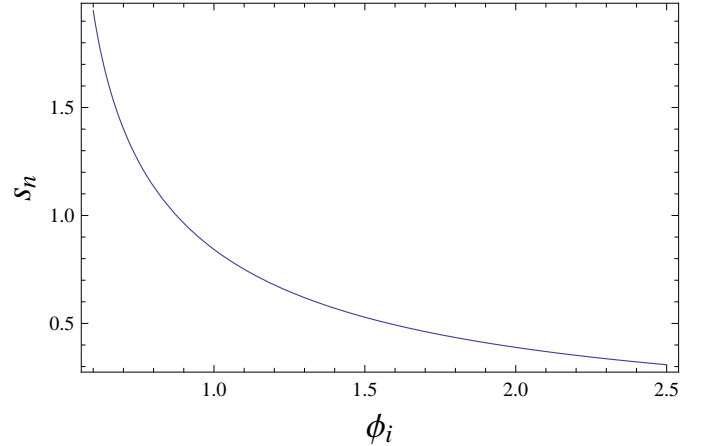}}
\caption{Dependence of $s_{n}$ as function of $\phi_{i}$ satisfying the flatness condition. $\Omega_{m0} =0.24$}
\end{figure}

\noindent
With the setup discussed in the previous section and assuming a flat FRW metric with scale factor $a(t)$, the Raychoudhury equation and the equation of motion for the axion field in the presence of non-relativistic matter ($p = 0$, $p$ being the pressure of the matter fluid) are given by,

\begin{eqnarray}
\frac{\ddot{a}}{a} = \frac{1}{3M_{pl}^{2}}(-{\dot{\phi}}^2+s\phi)-
\frac{\Omega_{m0}}{2a^{3}}H_{0}^2 \\
\ddot{\phi}+3\frac{\dot{a}}{a}\dot{\phi}+s = 0.
\end{eqnarray}

\noindent
Here $s = \frac{\mu^4}{f_{a}}$ and $\Omega_{m0}$ is the density parameter for matter fluid at present. Next we define the dimensionless quantities:

\begin{equation}
H_{0}t\rightarrow t_{n}\hspace{0.02mm}, \hspace*{4mm} \frac{\phi}{\sqrt{3}M_{pl}}\rightarrow \phi_{n}  ,\hspace*{4mm}  \frac{s}{\sqrt{3}M_{pl}H_{0}^2}\rightarrow s_{n}
\end{equation}

In terms of these dimensionless quantities, the equation (11) becomes,

\begin{equation}
 \frac{\ddot{a}}{a} = (-{\dot{\phi_{n}}}^2 +s_{n}\phi_{n})-\frac{\Omega_{m0}}{2a^{3}}
\end{equation}

Here the subscript `n' refers to the new quantities and the time derivative is taken with respect to $t_n$. The form of the equation (12) remains the same in terms of these new quantities. It is now straightforward to solve these two equations numerically given the initial conditions. For this we assume that the universe was matter dominated in the early time and the scalar field  was nearly frozen initially. This gives the following initial conditions:

\begin{equation}
 a(t_{i}) = \left(\frac{9\Omega_{0m}}{4}\right)^{1/3} t_{i}^{2/3},\hspace*{4mm} \dot{\phi_{n}}(t_{i}) =
0, \hspace*{4mm} \phi_{n}(t_{i}) = \phi_{i}.
\end{equation}

\noindent
With this one can now solve the system. We have three parameters, e.g $s$, $\phi_{i}$ and $\Omega_{m0}$. But these three parameters are not independent and are related by the condition $\Omega_{\phi0} = 1-\Omega_{m0}$, where $\Omega_{\phi0}$ is the density parameter for the scalar field at present. This condition arises due to our assumption of flat FRW spacetime (k=0). Hence essentially we have two parameters in our model. In Figure 1, we have shown the behaviour of $s_{n}$ as a function of $\phi_{i}$ satisfying the flatness condition.

Before studying the observational constraints on different model parameters, we show the behaviour of the equation of state for the axion field for different initial values of the field $\phi_{i}$. In figure 2 we show this behaviour. It is evident from the behaviour, that higher the initial values for the axion field, the behavior stays very close to the cosmological constant case $\omega = -1$. For smaller values of $\phi_{i}$, the equation of state for the axion field $\omega$ deviates significantly from $\omega = -1$ at present. We show the behaviour for $\Omega_{m0} = 0.24$. The conclusion is same for other values of $\Omega_{m0}$. We also show in Figure 3, the behaviour of the energy density for the axion field $\rho_{\phi}$ for different values of $\phi_{i}$ for $\Omega_{m0} = 0.24$. The $\rho_{\phi}$ remains nearly constant for higher $\phi_{i}$, whereas for lower values of $\phi_{i}$, although the $\rho_{\phi}$ is initially frozen, it decays sharply at later epochs. Here also, the overall behaviour remains same for other values of $\Omega_{m0}$.

\begin{figure}[t]
\psfrag{alphabet}{$a$}
\psfrag{w}{$\omega$}
\centerline{\epsfxsize=3.5truein\epsfbox{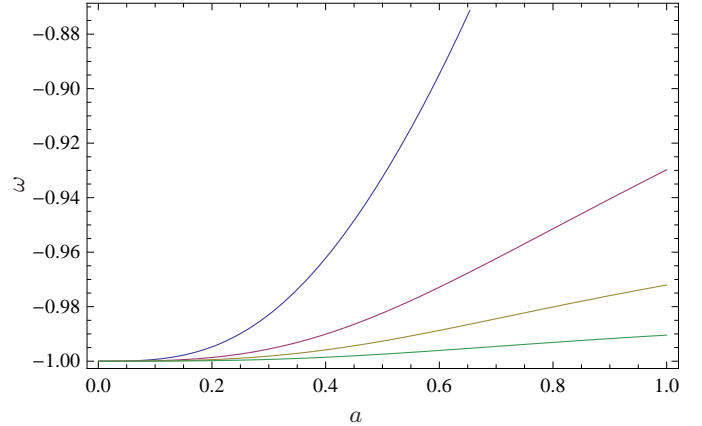}}
\caption{Evolution of the equation of state $\omega$ for the axion field as function of scale factor. From top to bottom, $\phi_{i} = 0.6,1,1.5, 2.5$. $\Omega_{m0} =0.24$}
\end{figure}

\section
{Observational Constraint On Model Parameters}

In this section, we constrain the parameters in the axion model with the assumption of a flat Universe by using the latest observational data including the Type 1a Supernovae Union2 compilation\cite{2010ApJ...716..712A}, the Baryon Acoustic Oscillation\cite{Eisenstein}  measurement from the SDSS\cite{BAO,BAO2}, the Cosmic Microwave background measurement given by WMAP\cite{WMAP7} observations and the H(z) data from HST key Project\cite{H0} . Here we limit ourselves to the background evolution of the universe.

\subsection{Type Ia Supernovae}

We consider the Supernovae Type Ia observation which is one of the direct probes for late time acceleration. It measures the apparent brightness of the Supernovae as observed by us which is related to the luminosity distance $d_{L}(z)$ 
defined  as 
\begin{equation}
d_{L}(z) = (1+z)\int_0^z\frac{dz^{\prime}}{H(z^{\prime})}\end{equation}

\begin{figure}[t]
\psfrag{rhophi}{$\rho_{\phi}$}
\psfrag{alphabet}{$a$}
\centerline{\epsfxsize=3.5truein\epsfbox{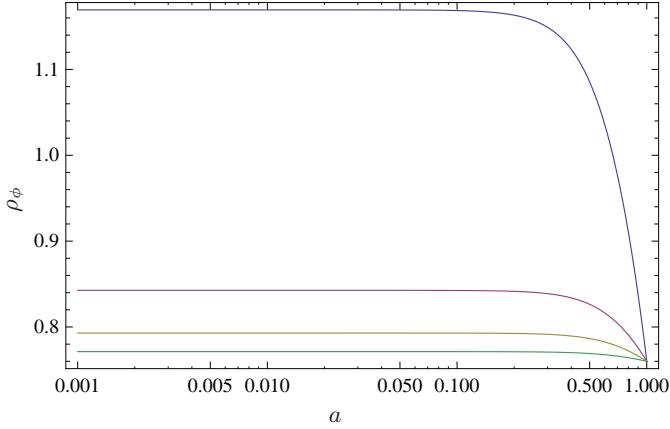}}
\caption{Behaviour of the energy density of the axion field $\rho_{\phi}$ for different values of $\phi_{i}$. From top to bottom, $\phi_{i} = 0.6,1,1.5,2.5$ respectively. $\Omega_{m0}=0.24$}
\end{figure}

With this we construct the distance modulus `$\mu$' which is experimentally measured
\begin{equation}
\mu = m-M = 5\log\frac{d_{L}}{Mpc}+25
\end{equation}
Where m and M are the apparent and absolute magnitudes of the Supernovae which are logarithmic measure of flux and luminosity respectively.

\subsection{Baryon Acoustic Oscillation}

\vspace{5mm}
Another observational probe that has been widely used in recent times to constrain dark energy models is related to the data from the Baryon Acoustic Oscillations measurements by the large scale galaxy survey.
In this case,one needs to calculate the parameter $D_{v}$ which is related to the angular diameter distance as follows
\begin{equation}
D_{v} = \left[\frac{z_{BAO}}{H(z_{BAO})}\left(\int_0^{z_{BAO}}\frac{dz}{H(z)}\right)^2\right] ^{1/3}.
\end{equation}
For BAO measurements we calculate the ratio 

{\Large$\frac{D_{v}(z = 0.35)}{D_{v}(z = 0.20)} $}.
 \vspace{1mm}
This ratio is a relatively model independent quantity and has a measured value $1.736 \pm 0.065$.

\subsection{Cosmic Microwave Background}
The CMB is sensitive to the distance to the decoupling epoch via the locations of peaks and troughs of the acoustic oscillations. We employ the ``WMAP distance priors''  given by the seven-year WMAP observations.This includes the``acoustic scale'' $l_{A}$,the `` shift parameter '' R and the redshift of the decoupling epoch of photons $z_{*}$
The acoustic scale $l_{A}$ describes the distance ratio $\frac{D_{A}(z_{*})}{r_{s}(z_{*})}$
\begin{equation}
l_{A} \equiv (1+z_{*})\frac{\pi D_{A}(z_{*})}{r_{s}(z_{*})}
\end{equation}
where $(1+z_{*})$ factor arises because $D_{A}(z_{*})$ is the proper angular diameter distance, where $r_{s}(z_{*})$ is the comoving sound horizon at $z_{*}$

We use the fitting function of $z_{*}$ proposed by Hu and Sugiyama
\begin{equation}
z_{*} = 1048[1+0.00124(\Omega_{b}h^2)^{-0.738}][1+g_{1}(\Omega_{m}h^2)^{g_{2}}]
\end{equation}
\begin{equation}
g_{1} = \frac{0.0783(\Omega_{b}h^2)^{-0.238}}{1+39.5(\Omega_{b}h^2)^{0.763}} , 
 g_{2} = \frac{0.560}{1+21.1(\Omega_{b}h^2)^{1.81}}
\end{equation}

The shift parameter R responsible for the distance ratio $\frac{D_{A}(z_{*})}{H^{-1}(z_{*})}$, given by
\begin{equation}
R(z_{*}) \equiv \sqrt{\Omega_{m0}H_{0}^{2}}(1+z_{*})D_{A}(z_{*}).
\end{equation}

The constraint on the shift parameter $R(z_{*})$ from the WMAP observations is quoted as $R(z_{*}) = 1.715\pm 0.021$. 

\subsection{H(z) Measurement}
Next we use  new determinations of the cosmic expansion history from red-envelope galaxies.  Stern et al. \cite{stern} have obtained a high-quality spectra with the Keck-LRIS spectrograph of red-envelope galaxies in 24 galaxy clusters in the redshift range $0.2 < z < 1.0$.  They complemented these Keck spectra with high-quality, publicly available archival spectra from the SPICES and VVDS surveys. With this, they presented 12 measurements of the Hubble parameter H(z) at different redshift. The measurement at $z=0$ was from HST Key project \cite{H0}.

\vspace{5mm}
Subsequently we use these four observational results to constrain our model parameters. As we mentioned earlier, we have three parameters to constrain, e.g,  $s_{n}, \phi_{i}$ and $\Omega_{m0}$ (or $\Omega_{\phi 0} = 1-\Omega_{m0}$). But these parameters are related by the flatness condition ($k = 0, \Omega_{m} + \Omega_{\phi} = 1$) and hence essentially we have two independent parameters to constrain.

To start with, we assume different values for the matter density parameter at present $\Omega_{m0}$ and put constraint on $\phi_{i}$. Once we obtain constraint on $\phi_{i}$, we shall use the flatness condition ($k=0, \Omega_{m} + \Omega_{\phi} = 1$) to obtain the corresponding constraint on $s_{n}$. As mentioned in the beginning, $s_{n}$ is related with various string related parameters in the following way:
\begin{equation}
s_{n} = \left(\frac{\mu^{4}}{f_{a}}\right)/\left(\sqrt{3}M_{pl} H_{0}^{2}\right).
\end{equation}

Hence constraining $s_{n}$ using observational data, one can put bounds on various other parameters of the model.
\begin{table*}
\begin{tabular}{|c|c|c|c|c|}
\hline
&$SN$&$SN+BAO$&$SN+BAO+CMB$&$SN+BAO+CMB+H(z)$\\
\cline{2-5}
&$\chi_{min}$= 543.336&$\chi_{min}$ = 544.46&$\chi_{min}$ = 544.649&$\chi_{min}$ =548.555\\
\cline{2-5}
$\Omega_{m0}=$&$\phi_{i}(bf)=1.033$, $s_{n}(bf)=0.81$&$\phi_{i}(bf)=1.08$, $s_{n}(bf)=0.76$&$\phi_{i}(bf)=1.26$, $s_{n}(bf)=0.64$&$\phi_{i}(bf)=1.22$, $s_{n}(bf)=0.66$\\
\cline{2-5}
 0.24&$\phi_{i}\geq 0.72$, $s_{n}\leq 1.32$ ($1\sigma$)&$\phi_{i}\geq 0.74$, $s_{n}\leq 1.29$ ($1\sigma$)&$\phi_{i}\geq0.78$, $s_{n}\leq 1.17$ ($1\sigma$)&$\phi_{i}\geq 0.78$, $s_{n}\leq 1.17$ ($1\sigma$)\\
 &$\phi_{i}\geq 0.63$, $s_{n}\leq 1.72$ ($2\sigma$)&$\phi_{i}\geq 0.64$, $s_{n}\leq 1.69$ ($2\sigma$)&$\phi_{i}\geq 0.66$, $s_{n}\leq 1.55$ ($2\sigma$)&$\phi_{i}\geq 0.66$, $s_{n}\leq 1.55$ ($2\sigma$)\\
\hline
$\Omega_{m0}=$&$\phi_{i}(bf)=2.5$, $s_{n}(bf)=0.29$&$\phi_{i}(bf)=2.5$, $s_{n}(bf)=0.29$&$\phi_{i}(bf)=2.5$, $s_{n}(bf)=0.29$&$\phi_{i}(bf)=2.5$, $s_{n}(bf)=0.29$\\
\cline{2-5}
 0.28&$\phi_{i}\geq 1.16$, $s_{n}\leq 0.66$ ($1\sigma$)&$\phi_{i}\geq 1.18$, $s_{n}\leq 0.65$ ($1\sigma$)&$\phi_{i}\geq 1.07$, $s_{n}\leq 0.72$ ($1\sigma$)&$\phi_{i}\geq 1.09$, $s_{n}\leq 0.7$ ($1\sigma$)\\
 &$\phi_{i}\geq 0.75$, $s_{n}\leq 1.13$ ($2\sigma$)&$\phi_{i}\geq 0.76$, $s_{n}\leq 1.12$ ($2\sigma$)&$\phi_{i}\geq 0.73$, $s_{n}\leq 1.2$ ($2\sigma$)&$\phi_{i}\geq 0.74$, $s_{n}\leq 1.17$ ($2\sigma$)\\
\hline
$\Omega_{m0}=$&$\phi_{i}(bf)=2.5$, $s_{n}(bf)=0.28$&$\phi_{i}(bf)=2.5$, $s_{n}(bf)=0.28$&$\phi_{i}(bf)=2.5$, $s_{n}(bf)=0.28$&$\phi_{i}(bf)=2.5$, $s_{n}(bf)=0.28$\\
\cline{2-5}
 0.3&$\phi_{i}\geq 1.29$, $s_{n}\leq 0.57$ ($1\sigma$)&$\phi_{i}\geq 1.31$, $s_{n}\leq 0.56$ ($1\sigma$)&$\phi_{i}\geq 1.18$, $s_{n}\leq 0.63$ ($1\sigma$)&$\phi_{i}\geq 1.21$, $s_{n}\leq 0.61$ ($1\sigma$)\\
 &$\phi_{i}\geq 0.81$, $s_{n}\leq 0.99$ ($2\sigma$)&$\phi_{i}\geq 0.82$, $s_{n}\leq 0.98$ ($2\sigma$)&$\phi_{i}\geq 0.76$, $s_{n}\leq 1.08$ ($2\sigma$)&$\phi_{i}\geq 0.78$, $s_{n}\leq 1.05$ ($2\sigma$)\\
\hline
$\Omega_{m0}=$&$\phi_{i}(bf)=2.5$, $s_{n}(bf)=0.27$&$\phi_{i}(bf)=2.5$, $s_{n}(bf)=0.28$&$\phi_{i}(bf)=2.5$, $s_{n}(bf)=0.28$&$\phi_{i}(bf)=2.5$, $s_{n}(bf)=0.28$\\
\cline{2-5}
 0.32&$\phi_{i}\geq 1.37$, $s_{n}\leq 0.52$ ($1\sigma$)&$\phi_{i}\geq 1.4$, $s_{n}\leq 0.51$ ($1\sigma$)&$\phi_{i}\geq 1.25$, $s_{n}\leq 0.57$ ($1\sigma$)&$\phi_{i}\geq 1.28$, $s_{n}\leq 0.55$ ($1\sigma$)\\
 &$\phi_{i}\geq 0.85$, $s_{n}\leq 0.9$ ($2\sigma$)&$\phi_{i}\geq 0.85$, $s_{n}\leq 0.89$ ($2\sigma$)&$\phi_{i}\geq 0.78$, $s_{n}\leq 0.99$ ($2\sigma$)&$\phi_{i}\geq 0.8$, $s_{n}\leq 0.96$ ($2\sigma$)\\
\hline
\end{tabular}
\caption{Constraints on $\phi_{i}$ and $s_{n}$ for various values of matter density parameter at present $\Omega_{m0}$. The ``bf'' denotes the best fit values.}
\end{table*}

\subsection{Results}

As mentioned above, we use the various observational data related to the background cosmology, mentioned in the previous section to constrain the parameter $\phi_{i}$ and subsequently the parameter $s_{n}$. The bounds obtained on different parameters are shown in the Table 1. The main result is that we  get a lower bound on $\phi_{i}$ for different values of the present day  matter density parameter, $\Omega_{m0}$. This is consistent with the behaviour of the equation of state $\omega$ that we show in Figure 2. For higher values of $\phi_{i}$ the cosmic evolution is very close to the cosmological constant whereas for  lower values, one gets more deviation from the cosmological constant. Hence a lower bound on $\phi_{i}$ actually constrains the deviation from the cosmological constant. For the parameter $s_{n}$ we get an upper bound. Moreover, using the bounds on $\phi_{i}$ as well as the definitions (13) and (15), it is easy to check that the constrained value for $\phi_{i}$ is larger than $M_{pl}$. This is an important criteria  for this model to be viable.  

Next, denoting the constrained value of $s_{n}$ by $s_{nB}$ which is tabulated in Table 1, one can write using (13), (10) and definition of $s_{n}$ (mentioned after equation (12)):

\begin{equation}
\sqrt{3}M_{PL}H_{0}^{2}s_{nB} = \frac{\mu_{1}}{f_{a}}+\frac{\mu_{2}}{f_{a}}
\end{equation}
where $\mu_{1}$ and $\mu_{2}$ are  defined after equation (10).  To start with, we assume the dominant contribution comes from the second term. Hence dropping $\mu_{1}$ in the above expession, one can now write:
\begin{equation}
\frac{\mu_{2}}{f_{a}} = \sqrt{3}M_{PL}H_{0}^{2}s_{nB}.
\end{equation}
Using the expression for $\mu_{2}$ defined after equation (10), we can now write
\begin{equation}
\sqrt{2}c M_{SB}^{4}\frac{e^{2A_{0}}}{g_{s}}\left(\frac{R^{2}}{\alpha'L^{2}} \right) = M_{PL}^{2}H_{0}^{2}s_{nB}.
\end{equation}
Assuming $H_{0} \sim 10^{-42}$Gev and neglecting the dependence of $c$, we get the final expression as:
\begin{equation}
M_{SB}\approx10^{-30}M_{PL}\left[\frac{e^{-2A_{0}}}{\beta} \right]^{1/4} s_{nB} 
\end{equation}
where, $\beta$ = $\frac{R^{2}}{g_{s}\alpha'L^{2}}>1$. Now if we assume $M_{SB} = \alpha$ TeV, where $\alpha$ is a number determining the SUSY breaking scale, then
\begin{equation}
\left[\frac{e^{-2A_{0}}}{\beta} \right] \approx \alpha^{4}10^{56}(s_{nB})^{-4}.
\end{equation}
Here $s_{nB}$ is the observational bound on the the parameter $s_{n}$ mentioned in the Table 1. From the above expression, either one can assume a value for the combination $\left[\frac{e^{-2A_{0}}}{\beta} \right]$ which is consistent with string theory, and get the bound on $\alpha$, the SUSY breaking scale, or assume a SUSY breaking scale $\alpha$ and get a bound on the combination $\left[\frac{e^{-2A_{0}}}{\beta} \right]$. In figure 4, we show this bound at $2\sigma$ confidence level.

\begin{figure}[t]
\psfrag{a}{$\alpha$}
\psfrag{b}{$\log_{e} [\frac{e^{-2A_{0}}}{\beta}]$}
\centerline{\epsfxsize=3.5truein\epsfbox{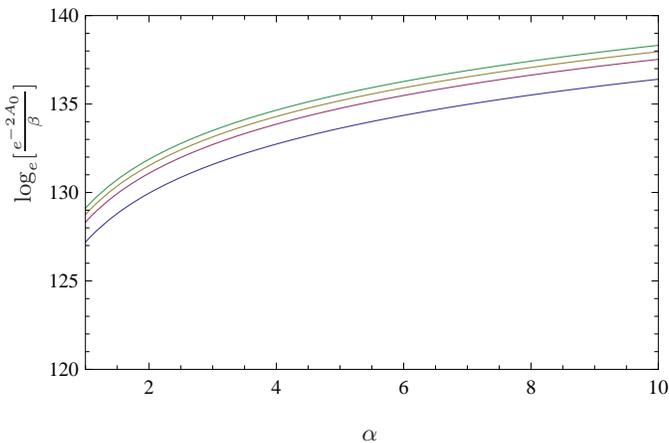}}
\caption{The allowed values for $\left[\frac{e^{-2A_{0}}}{\beta} \right] $  at $2\sigma$ confidence level for various SUSY breaking scales denoted by $\alpha$. This is  for SN+BAO+CMB+H(z) observations. $\Omega_{m0}=0.24,0.28,0.3,0.32$ from bottom to top.}
\end{figure}

\begin{figure}[t]
\centerline{\epsfxsize=3.5truein\epsfbox{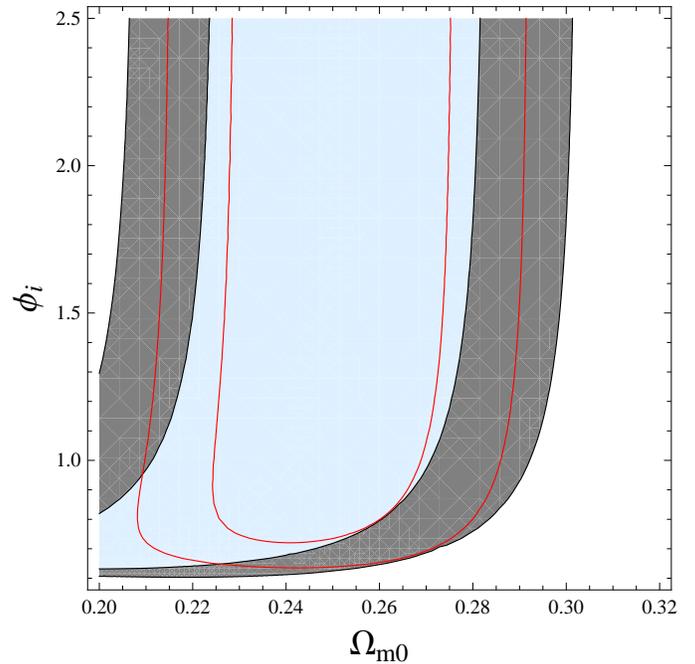}}
\caption{The $1\sigma$ and $2\sigma$ contours in the $\phi_{i}-\Omega_{mo}$ plane. The shaded contours are for SN+BAO whereas the solid lines are for SN+BAO+CMB+H(Z) data.}
\end{figure}

\begin{figure}[t]
\centerline{\epsfxsize=3.5truein\epsfbox{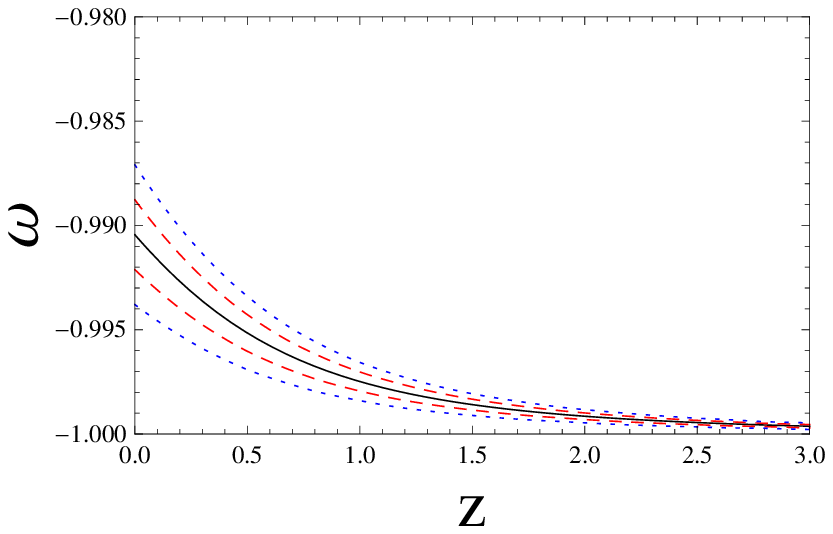}}
\caption{The reconstructed equation of state $\omega$ as a function of $z$ at $1\sigma$ and $2\sigma$ confidence level. we take SN+BAO+CMB+H(Z) data. The solid line is for the best fit values, whereas the dashed and dotted lines are for $1\sigma$ and $2\sigma$ confidence level. }
\end{figure}

Next we redo the analysis by keeping $\Omega_{m0}$ as a free parameter. For this, we assume a range for $\Omega_{m0}$ between the values $0.2$ and $0.32$. In figure 5, we show the allowed region in the $\phi_{i}-\Omega_{m0}$ plane. It is evident that there is a lower bound on $\phi_{i}$ which controls the deviation from the $\Lambda$CDM behaviour. In figure 6, we show the reconstructed equation of state.  It is clear that even at the $2-\sigma$ confidence level, the allowed equation of state is extremely close to the cosmological constant. Hence  allowed model for all practical purposes will behave like cosmological constant.  This is unlike other dark energy models,  where although cosmological constant is always consistent, the data still allows a significant deviation from $\Lambda$CDM. But in our case, data only allows the model to behave very close to $\Lambda$CDM.

\section{Conclusion}

We investigate the observational bounds on the recently proposed axions models in string theory which can be a  suitable candidate for dark energy. The evolution of the universe is controlled by two parameters: one is $s_{n}$ which is a combination of various stringy parameters and the other one is $\phi_{i}$, the initial value of the axion field. In addition the present energy density of the matter $\Omega_{m0}$ also affects the evolution. But these three parameters are not completely independent as they can be related by imposing the flatness condition on the space part of the universe. The parameter $\phi_{i}$ controls the deviation of the model from the $\Lambda$CDM behaviour.  One interesting aspect of our investigation is that one can use cosmological observations to put constrain on the SUSY breaking scale fixing  various stringy parameter. Alternatively, if one fixes the SUSY breaking scale, one can put constraints on the combination of various stringy parameter. Also our study shows that the allowed behaviour of the model is extremely close to the $\Lambda$CDM behaviour making it indistinguishable from $\Lambda$CDM for all practical purposes.

\section{Acknoweldgement}    
GG acknowledges the financial support from the CSIR, Govt. of India through the Senior Research Fellowship. AAS acknowledge the financial support from the SERC, DST, Govt. of India through the research grant DST-SR/S2/HEP-043/2009.

\end{document}